# Silicide Induced Surface Defects in FePt nanoParticle *fcc*-to-*fct* Thermally Activated Phase Transition

Shu Chen,[a] Stephen L. Lee,[a] Pascal André [a,b,c*]

[a] School of Physics and Astronomy, SUPA, University of St Andrews, St Andrews KY16 9SS, UK
[b] RIKEN, Wako 351-0198, Japan
[c] Department of Physics, CNRS-Ewha international Research Center (CERC), Ewha W. University, Seoul 120-750, Republic of Korea

Magnetic nanoparticles (MnPs) are relevant to a wide range of applications including high density information storage and magnetic resonance imaging to name but a few. Among the materials available to prepare MnPs, FePt is attracting growing attention. However, to harvest the strongest magnetic properties of FePt MnPs, a thermal annealing is often required to convert face-centered cubic as synthesized nPs into its tetragonal phase. Rarely addressed are the potential side effects of such treatments on the magnetic properties. In this study, we focus on the impact of silica shells often used in strategies aiming at overcoming MnP coalescence during the thermal annealing. While we show that this shell does prevent sintering, and that *fcc*-to-*fct* conversion does occur, we also reveal the formation of silicide, which can prevent the stronger magnetic properties of *fct*-FePt MnPs from being fully realised. This report therefore sheds lights on poorly investigated and understood interfacial phenomena occurring during the thermal annealing of MnPs and, by doing so, also highlights the benefits of developing new strategies to avoid silicide formation.

## I. Introduction

Nanoparticles (nPs) have been the focus of intense research activities in terms of preparation and application of a wide range of materials. While magnetic nPs are part of a subset overwhelmingly dominated by iron oxide derivatives [1-11], iron platinum (FePt) nPs have attracted a lot of attention both experimentally and computationally, which has arisen because of their magnetic properties and potential use for both information storage devices and biomedical applications [12-18]. Especially noteworthy are the high magnetic anisotropy of the order of $10^6$ erg/cm$^3$ at room temperature [19,20], and being a building block for multi-functional materials [13-15], as well as having both a large chemical stability and strong MRI contrast agent properties suitable for in-vivo imaging [17]. Nonetheless, a key challenge associated with this material is that most solution based syntheses tend to lead to a face centered cubic (*fcc*) crystalline phase, whereas the strongest magnetic properties are associated with the tetragonal (*fct*) phase, which is challenging to synthesize directly [21-30]. As a consequence, various strategies are being investigated to convert as-prepared *fcc*-FePt nPs into their *fct* phase. Whilst straightforward at first sight, thermal annealing under reducing atmosphere tends to lead to coalescence [31] unless the nPs are encapsulated in inorganic matrices or shells.

Aside from the salt based strategy [32-35], silicon oxide presents a versatile chemistry suitable for the attachment of chromophores to the silicon atoms [36-40], as well as the growth most often by sol-gel technique of silica shells around a range of metallic, semiconductors or magnetic nPs [41-44]. The use of silica shells has therefore allowed the development of multifunctional nPs [45], soluble temperature nanoprobes [46], drug delivery [47], biocatalysts [48], as well as *fcc*-to-*fct* phase conversion [49-51], although very few reports have commented directly or indirectly on the possible detrimental side-effects of such strategies [31,52,53]. Beyond the need of high temperature post-treatments and subtle chemistry to minimise the number of empty silica shells, the coating and removal of the silica shell can impact on the FePt magnetic nPs.

On a more general note, it is known that with their high surface to volume ratio, alteration of the nP interface can lead to a large effect on their properties [54]. For instance, a variation of saturation magnetization, coercitivity or blocking temperature can be observed when coating iron oxide nPs with different organic molecules [5], or FePt nPs with a gold layer [55,56]. These are often rationalised in terms of magnetic dead-layers [2,8,57,58]. In the present study, we report on little investigated interfacial side effects of the FePt-SiO$_2$ core-shell strategy used to implement *fcc*-to-*fct* crystalline phase conversion. The colloidal nPs were prepared by solution chemistry and their morphological, crystalline and magnetic properties characterised. Whilst we confirmed that silica shells efficiently prevent sintering during the thermal annealing, and that *fcc*-to-*fct* conversion occurred, we also provide evidence for the formation of silicide, which strongly altered the magnetic properties of the nPs. As this prevents the stronger magnetic properties of the obtained *fct*-FePT nPs from being fully realised, we discuss the formation of the silicide which can contribute to explain FePt nP property variations reported in the literature, and consequently points at alternative pathways aiming at overcoming this interfacial issue.

## II. Experimental Section

### II.1. FePt nPs preparation

**Materials:** All chemical reagents, unless otherwise stated, were purchased from Sigma, used without further purification but degassed before use: Disodium tetracarbonylferrate-dioxane complex (Na$_2$Fe(CO)$_4$·1.5 C$_4$H$_8$O$_2$), platinum(II) acetylacetonate (Pt(acac)$_2$, 97 %), oleylamine (70 %), oleic acid (90 %), tetra-ethylorthosilicate (TEOS, Si(OC$_2$H$_5$)$_4$, ≥ 99.0 %), (3-aminopropyl) triethoxy silane (APTES, H$_2$N(CH$_2$)$_3$Si(OC$_2$H$_5$)$_3$, ≥ 98 %), IGEPAL® CO-520 (NP-5, (C$_2$H$_4$O)$_n$·C$_{15}$H$_{24}$O, n ~ 5). When available ACS grade solvents were selected: dibenzyl ether (≥ 98.0 %), hexane (99.0 %), ethanol (≥ 99.5 %), chloroform (≥ 99 %, anhydrous), methanol (≥ 99.8 %), cyclohexane (GC, ≥ 98.0

__________
* Corresponding author. Present address: Department of Physics, CERC, Ewha W. University, Seoul 120-750, Republic of Korea ; RIKEN, Wako 351-0198, Japan
*E-mail address:* pjpandre@riken.jp (P. André)

%), ammonium hydroxide aqueous solution (NH$_4$OH, 28.0 to 30.0 %).

**FePt synthesis:** All the nanoparticle syntheses were carried out inside a glove box. In a typical synthesis a mixture of Pt(acac)$_2$ (1 mmol), oleyl amine (8 mmol) and oleic acid (4 mmol) in 10 mL of dibenzyl ether were transferred in a 50 mL round bottom flask connected to a condenser. Under stirring, the mixture was heated up to 100 °C for 1 h to remove oxygen and moisture, then Na$_2$Fe(CO)$_4$ (1 mmol) in 10 mL dibenzyl ether mixture was added and the mixture was heated up to 150 °C for 1 h. The temperature was then further increased to maintain the reflux at ~ 300 °C for 3 h. The resulting dark solution was cooled down to room temperature and after washing with hexane and ethanol, the product was collected by centrifugation and could easily be redispersed in solvent such as hexane and chloroform [17,50].

*fcc*-FePt-silica: The *fcc*-FePt-silica nPs were prepared by hydrolysis of tetraethylorthosilicate (TEOS) and the silica shell surface was functionalized with (3-aminopropyl) triethoxysilane [42,46]. Reverse microemulsions were prepared under vigorous stirring by mixing 10 mL of cyclohexane, 1.3 mL of poly oxyethylene nonylphenyl ether non-ionic surfactant (NP-5) and 50 μL of DI H$_2$O. 2 mg (~ 3-4 nmol) of FePt nPs were then dispersed in 1 mL cyclohexane and added dropwise into the reverse microemulsion. After 15 min, 80 μL TEOS was added dropwise. After another 15 min, 150 μL NH$_4$.H$_2$O (28-30 %) was added dropwise. The solution was kept under constant stirring at room temperature for 72 h. To form amine functionalized FePt-silica nPs, 100 μL of APTES was added after 48 h and kept stirring for another 24 h. The nPs were precipitated by centrifugation after addition of 3 mL of ethanol and 2 mL of methanol. The nPs were redispersed in 5 mL of ethanol and precipitated by centrifugation after addition of 10 mL of hexane. This step was repeated up to 6 times to completely remove the surfactant. FePt-silica nPs were stable both in ethanol and DI water.

*fct*-**FePt nPs:** The *fct* phase FePt nPs were obtained by annealing *fcc*-FePt-silica nPs deposited into a quartz boat sitting into a glass tube in order to allow control of the atmosphere. The furnace was set to 700 °C for 4 h with N$_2$/H$_2$ (9:1) gas flowing through the setup. After cooling back to room temperature, the silica shell was dissolved with concentrated NaOH solution and functionalized with cysteamine to form *fct*-FePt nPs. Experimentally, 16 mL of 4 M NaOH water solution contains 1.2 g of cysteamine was added into fine milled with 170 mg *fct*-FePt-silica nPs. The mixture was kept under inert atmosphere and vigorous stirring for 48 h to both dissolve the silica shell and functionalize the nP surface with cysteamine. After addition of DI H$_2$O and ethanol, the nPs were collected by centrifugation.

### II.2. FePt nPs characterization

**TEM:** Transmission electron microscopy (TEM) images were recorded using a Gatan CCD camera on a JEOL JEM-2011 electron microscope operating at 200 kV.

The chemical composition of FePt nPs was examined by energy-dispersive X-ray spectroscopy (EDX) using alternatively an Oxford Link system installed on a JEM-2011 microscope, and an Oxford Inca system integrated in a JSM-5600 scanning electron microscope.

**XRD:** Wide-angle powder X-ray diffraction (XRD) data were collected on a Stoe STADI/P powder diffractometer operating in transmission mode and with a small angle position sensitive detector. The instrumental line broadening was subtracted from every spectrum. The crystalline size of the nanoparticles was calculated with the Scherrer equation [59]:

$$D_{XRD} = \frac{0.9\lambda}{B\cos\theta} \qquad (eq.\ 1)$$

where $D_{XRD}$ is the "average" dimension of the crystallites, $\lambda$ is the wavelength of the X-ray source (for Fe$_{K\alpha 1}$ source, $\lambda$ is equal to 0.193604 nm), $B$ is the full width at half maximum of the peak intensity, $\theta$ is the glancing angle.

The atomic composition, $x$, of Fe$_x$Pt$_{1-x}$ nPs was calculated based on FePt lattice constant vs. composition linear curves reported by Bonakdarpour *et al.* [60]:

| $x_{\%Fe} < 40\ \%$ | $a = -0.0014\ x_{\%Fe} + 3.929$ | (eq. 2) |
| $x_{\%Fe} > 40\ \%$ | $a = -0.0041\ x_{\%Fe} + 4.039$ | (eq. 3) |

**SQUID:** A 5.0 Tesla Superconducting Quantum Interference Device (SQUID) from Quantum Design (MPMS XL$^{TM}$) was used to characterize the nPs magnetic properties. The nPs were dispersed in a polyvinylpyrrolidone matrix (V$_{Polymer}$/V$_{nPs}$ = 20) to prevent interactions between the nPs, and the resulting sample was loaded into low background gelatin capsules. Zero-Field Cooled and Field Cooled (ZFC/FC) measurements were completed as follow: the sample was first cooled from room temperature to 2 K without any external field, next a small field 100 Oe was applied and the nPs magnetization was recorded as the temperature was increased up 275 K. The FC curve was obtained by cooling the sample back to 2 K under a 100 Oe magnetic field. The magnetization was then measured while the temperature was increased up to 275 K. Hysteresis measurements were completed at temperatures of 2 K and 300 K. The background magnetization including the gelatine capsules and the PVP matrix was subsequently subtracted.

### III. Results & Discussion

FePt nPs were synthesized by thermal decomposition as described in the experimental section. Figure 1a presents TEM images of various magnifications (a1 to a3) of the product obtained after extraction from the synthetic media. The main characteristics of the nPs are summarized in Table 1. The nPs population (Figure 1a1 and 2) is homogenous and composed of nPs of a sized of about 6 nm, which is consistent with the literature [17]. These nPs were then dispersed in micelles based on a non-ionic surfactant, cyclohexane and water, a ternary colloidal system, which is used to control chemical syntheses of a wide range of nanomaterials [26,50,61-68]. In the present case, they were used to coat the FePt nPs with a silica shell and TEM images of the resulting core-shell structures are shown in Figure 1b. The ratio of nanoparticles and surfactant was balanced so that no empty silica nP could be observed on the TEM grids, while also minimizing the amount of silica beads filled with multiple FePt nPs. The amount of TEOS precursor was then adjusted to lead to hybrid nPs presenting a total diameter of about 40 nm, while between one and three FePt nPs can be found in each silica shell. This coating process altered neither the shape (Figure 1a3 and b3) nor the crystallinity of the FePt cores.

**Table 1.** XRD, and TEM-EDX data of the FePt nPs: crystalline grain size of the nPs ($D_{XRD}$), Fe/Pt ratio ("x" in Fe$_x$Pt$_{1-x}$), elemental composition based on TEM-EDX (%X).

| Sample | *fcc*-FePt | *fcc*-FePt-silica | *fct*-FePt |
|---|---|---|---|
| $D_{XRD}$ (nm) | 3.6 ± 0.1 | 3.6 ± 0.1 | 6.1 ± 0.1 |
| $D_{TEM}$ (nm) | 5.8 ± 0.8 | -- | 8.0 ± 3.1 |
| Fe$_x$Pt$_{1-x}$ (%) | 43 ± 1[‡] / 45 ± 1[*] | 44 ± 1[‡] | 42 ± 2[*] |
| %Fe | 45.5 ± 0.5 | -- | < 39 ± 4 |
| %Pt | 54.5 ± 0.5 | -- | < 52 ± 4 |
| %Si | -- | -- | > 9 ± 4 |

[¶] 4 h @ 700 °C, [‡] deduced from XRD, [*] deduced from TEM-EDX.

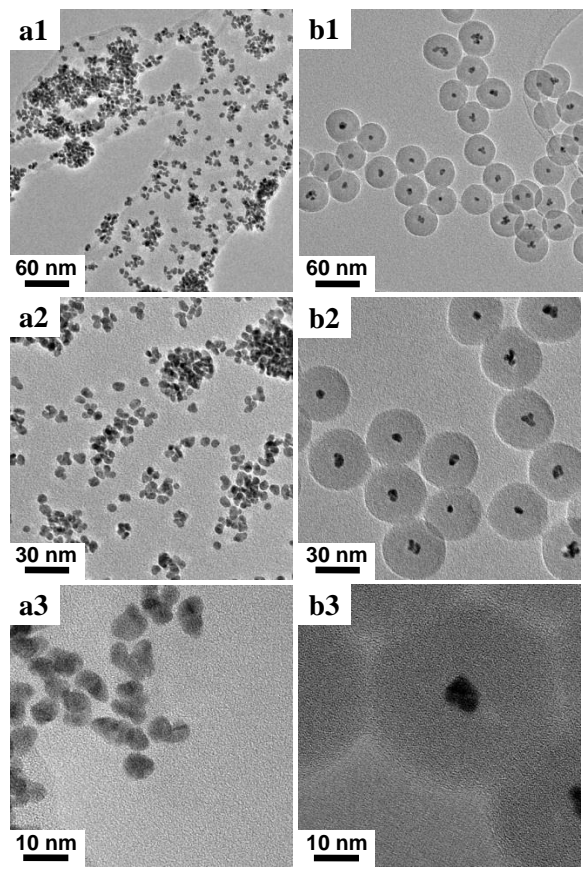

**Figure 1.** TEM images of *fcc*-FePt nPs (**a**), *fcc*-FePt-silica nPs (**b**).

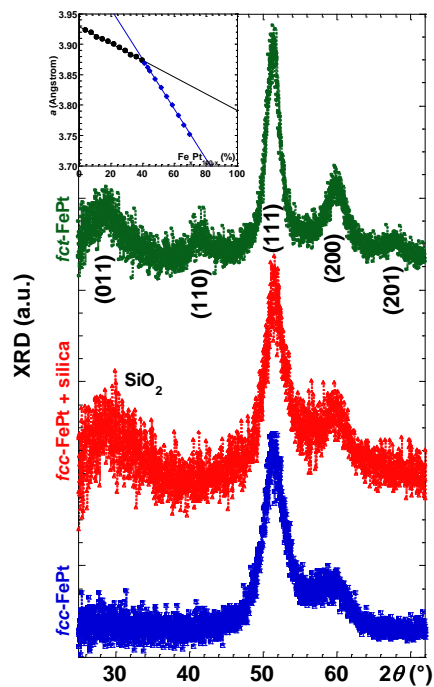

**Figure 2.** Background subtracted and normalized XRD of *fct*-FePt nPs annealed at 700 °C for 4 h (top green curve, ♦), *fcc*-FePt-silica nPs (middle red curve, ▲) and *fcc*-FePt nPs (bottom blue curve, ■). Insert: XRD based FePt lattice constant vs. composition.[60]

Figure 2 shows the background subtracted and normalized XRD data of the nPs. These spectra are characteristic of the *fcc*-phase of FePt (JCPDS # 29-0717) with the (111) and (200) XRD peaks clearly visible around 51° and 60°, respectively. As shown by Bonakdarkpour *et al.*, the lattice parameter is related to the FePt alloy composition (see insert of Figure 2). The linear variation (eq. 2 and 3) displayed on this calibration curve was then used to gain insight into the nPs composition for both as-synthesized and silica coated FePt nPs. As shown in Table 1, the nPs were found to contain about 43-44 % of iron. Within the experimental precision, this composition was found to be independent of the nPs coating and it was also found to be consistent with EDX measurements (Table 1 and Figure 3a). For the silica-coated nPs, a broad XRD peak characteristic of $SiO_2$ can be observed at ~28° (Figure 2). The crystalline grain size, $D_{XRD}$, was found to be ~3.6 nm for both *fcc*-FePt and *fcc*-FePt-silica nPs (Table 1), which is smaller but in relatively good agreement with the TEM data presented in Figure 1.

To convert the nanoparticles from *fcc* to *fct* phase, a thermal annealing of 4 h was completed under inert and reducing atmosphere. As illustrated in Figure 4a, the silica shells were found to partially fuse (4a1 and 4a2) but nonetheless to leave the FePt nPs rather homogenously dispersed within the matrix. This illustrates that silica shells do prevent sintering of the nPs during the annealing process. High magnification TEM images (Figure 4a3 and a4) display noticeable features including *i*) more spherical, *ii*) slightly larger (~10 nm) and *iii*) somewhat more polydispersed nanocrystals than the initial bean-shaped *fcc*-FePt nPs (Figure 1a).

The silica matrix was then dissolved in a concentrated sodium hydroxide solution, which also contained cysteamine. This molecule was used to prevent the nP aggregation by functionalising them as the thiol groups bond to the surface of the nPs and as the amine groups make the nPs water soluble. The resulting nPs are shown in Figure 4b. These TEM images confirm that whilst the silica matrix efficiently prevented sintering of the nPs during the annealing step, these nPs nonetheless appear larger than those that were initially encapsulated. This could naturally results in some instances from having several nPs per silica beads as can be seen in Figure 1b, and from the partial melting of the silica coating mentioned earlier (Figure 4a). High magnification TEM images (Figure 4b3 to b5) show nPs which tend to aggregate and which present high crystallinity, as illustrated by the presence of the facets and lattice fringes at the surface of the nPs. Fourrier transforms of these lattice fringes were used to obtain a diffraction pattern, which is presented in Figure 4b5. There resulting d-spacing of ~ 0.222 ± 0.005 nm is consistent with the (111) miller indices.

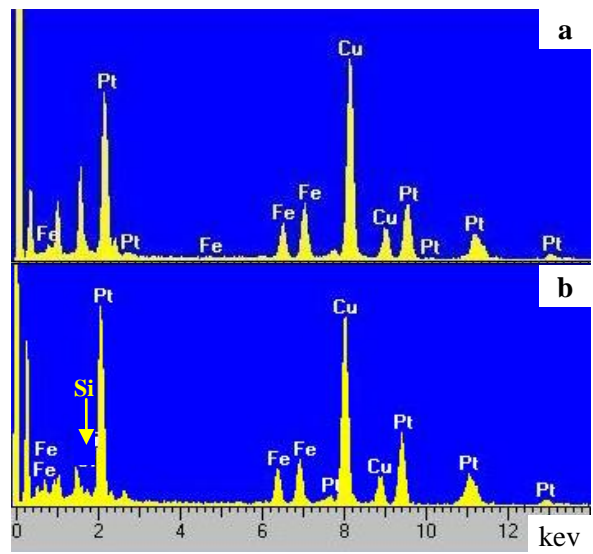

**Figure 3.** EDX data of *fcc*-FePt nPs (**a**) and *fct*-FePt nPs annealed at 700 °C for 4 h (**b**).

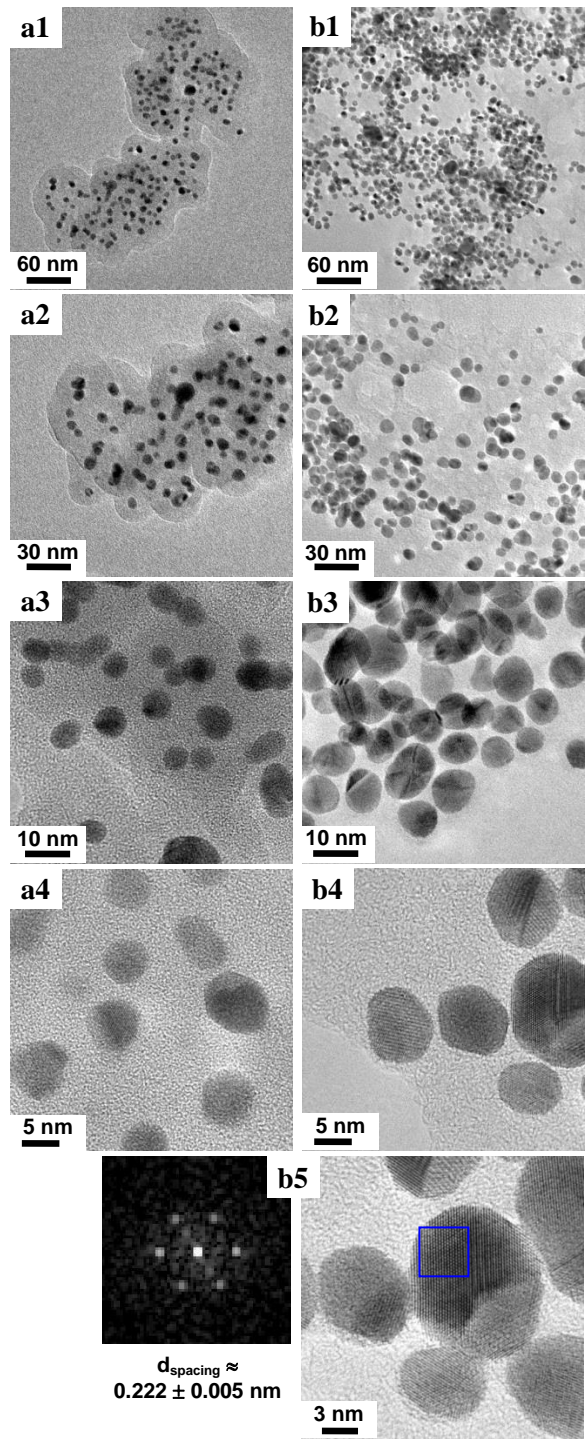

**Figure 4.** TEM images *fct*-FePt-silica nPs (**a**) and *fct*-FePt nPs (**b**), both after annealing at 700 °C for 4 h.

As illustrated in Figure 2, the *fct* crystal phase is evidenced after annealing by the XRD peaks around 29°, 41° and 68°, corresponding to the (001), (110) and (201) ordering reflections, respectively (JCPDS # 43-1359). From the Scherrer equation (eq. 1), the crystalline size of the annealed *fct*-FePt nPs was found to be about ~ 6 nm, larger than the initial *fcc*-FePt nPs but nonetheless smaller than the TEM size (Table 1) as observed before annealing.

The *fct* ordering parameter was then calculated with the intensity ratio of the (110) and (111) peaks [69],

$$S^2 = \frac{\{I_{(110)} / I_{(111)}\}_{obs}}{\{I_{(110)} / I_{(111)}\}_{bulk}} \quad (eq.\ 4)$$

In the above equation, $S$ is the chemical ordering parameter, which is widely used to characterized the *fcc* ($S = 0.0$) to *fct* ($S = 1.0$) phase transition. $I$ is the intensity of the XRD diffraction peaks, $\{I_{(110)}/I_{(111)}\}_{obs}$ is the ratio of the (110) and (111) XRD peak intensities experimentally measured, which in the present case is about ~ 0.12. This value has to be compared with the $\{I_{(110)}/I_{(111)}\}_{bulk}$ which is about ~ 0.27 as obtained from the PDF library card 03-065-9121. The resulting *fct* order parameter is then of ~ 0.7.

The magnetic characterisation of the three sets of nanoparticles was then completed to quantify the impact of the phase conversion. The main magnetic properties extracted from Figure 5 are summarised in Table 2. The left hand-side column of Figure 5 displays the zero-field-cooled (ZFC) and field-cooled (FC) measurements, each of them represented by the black and the blue curves, respectively. *fcc*-FePt nPs presented a blocking temperature $T_b$ of about ~ 90 K, corresponding to the maximum of the ZFC curves and signalled by a dashed line in Figure 5. The silica coating results in a drastic decrease of $T_b$ down to 50 K. This is nonetheless consistent with the literature for FePt nPs [41] and associated with the bonding of the oxide at the surface of the magnetic nPs. In contrast, after annealing and removal of the silica shell, the fct-FePt nPs showed a slight increase of $T_b$ up to 80 K. The relatively small amplitude of the recovery is, however, somehow surprising considering that fct-FePt is known to be strongly magnetic. Despite having only reached a 70 % conversion to fct phase, a blocking temperature above room temperature would have been expected for this nP range of size [29,34]. Reporting and discussing this apparent discrepancy is the main goal of this manuscript.

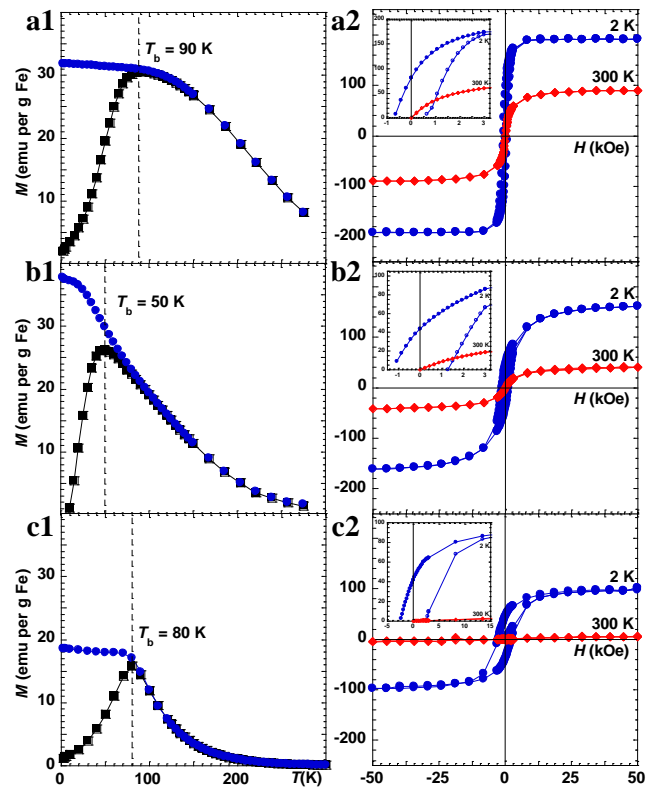

**Figure 5.** ZFC-FC under a 100 Oe magnetic field (**1**) and hysteresis curves at 2 K and 300 K (**2**) of *fcc*-FePt nPs (**a**), *fcc*-FePt-silica nPs (**b**) and *fct*-FePt nPs annealed at 700 °C for 4 h (**c**). The dashed line materializes the blocking temperature.

**Table 2.** SQUID data of the FePt based nPs: blocking temperature ($T_b$), magnetic coercivity ($H_c$), remanant magnetisation ($M_r$), magnetisation at saturation ($M_s$), saturation to remanent magnetisation ratio ($M_s/M_r$), magnetocrystalline anisotropy ($K_u$), and the dead layer per centage (%$t_{dl}$).

| Sample | fcc-FePt | fcc-FePt-silica | fct-FePt |
|---|---|---|---|
| $T_b$ (K) | 90 | 50 | 80 |
| $H_c$ @2K (kOe) | 0.7 | 1.3 | 2.5 |
| $M_r$ @2 K (emu/$g_{Fe}$) | 81.7 ± 0.1 | 43.7 ± 0.1 | 41.7 ± 0.2 |
| $M_s$ @2 K (emu/$g_{Fe}$) | 192 ± 1 | 145 ± 3 | 91 ± 3 |
| $M_s/M_r$ @2 K | 2.35 ± 0.02 | 3.32 ± 0.08 | 2.18 ± 0.08 |
| $K_u$ @2 K (μeV/nm$^3$) | 77 ± 6 | 215 ± 12 | 146 ± 20 |
| $M_s$ @300 K (emu/$g_{Fe}$) | 86 ± 1 | 35 ± 1 | 3 ± 1 |
| %$t_{dl}$ @300 K (nm) | 24.8 ± 0.1 | 29.9 ± 0.1 | 33.1 ± 0.1 |

The right hand-side of Figure 5 presents hysteresis loops of the magnetisation as a function of the applied magnetic field completed at temperatures of 2 K and 300 K. The insets of these figures expand parts of the magnetisation curves in order to highlight the hysteretic regions. The nPs of all three sets of nP type considered in this work appear to be in the superparamagnetic regime at 300 K as no hysteresis could be identified in contrast of the characterisation completed at 2 K. Both results are consistent with the blocking temperature measured with the ZFC-FC curves.

At 2 K, the coercivity of the nPs particles almost double after having coated the magnetic core with a silica shell (inserts in Figure 5a2 and b2) ; $H_c$ almost doubles again once 70 % of the nPs has been converted into the fct phase (insert in Figure 5c2) to reach 2.5 kOe. This confirms the fcc to fct crystal phase conversion. Simultaneously, the remanent magnetisation, $M_r$, decreases once the fcc-FePt nPs have been coated with a silica shell, and it only preserves its value once the phase conversion has been completed.

Well below the blocking temperature, the magnetization at saturation, $M_s$, and the magnetocrystalline anisotropy, $K_u$, values were obtained for each sample from the fits of the high field magnetisation curves with the saturation approach law as described below [70-74]:

$$M(H) = M_s \left[1 - b.\left(\frac{K}{M_s H}\right)^2\right] + \chi H \quad \text{(eq. 5)}$$

with $b = 8/105$ the coefficient associated with cubic anisotropy materials, and $\chi$ the high magnetic field susceptibility, which was found to be negligible in the present work. $K_u$ was found to be a few orders of magnitude lower than the value associated with fct-FePt bulk materials, which is ~60 meV/nm$^3$. This is consistent with the value reported in the literature for measurements completed at 5 K on similarly prepared nPs [49]. Furthermore, we note that $K_u$ is influenced significantly by factors such as temperature [75,76], dipolar interactions [37], nP stoichiometry [20,77], chemical ordering and surface [20,77-81], as well as size [82,83]. The last three factors suggest that different chemical pathways could lead to different $K_u$ values. In the present case, the fct-FePt nPs present a magnetocrystalline anisotropy, which is twice as large as the value associated with fcc-FePt nPs.

Finally, both silica coating and annealing induced noticeable drops in magnetization at saturation. This results likely from the fcc-FePt nPs surface alteration by the silane, which is consistent with observations reported in a range of magnetic nanoparticles whether they were coated with silica, metal or organic molecules. In the present case, the 25 % decrease of $M_s$ suggests defect formation once the oleic acid and oleyl amine have been replaced by silane bonds, which lead to the higher magnetocrystalline anisotropy of the fcc-FePt-silica nPs when compared with fcc-FePt nPs. In contrast and even if the fct fraction is only of the order of ~ 0.7, the further decrease of the fct-FePt magnetisation after thermal annealing does not match other reports suggesting much higher values [49]. We attributed this alteration of the magnetic nPs to silicide formation at the surface of the nPs as suggested by TEM-EDX data. Figure 3b displays an EDX peak at 1.7474 keV, which is characteristic of silicon (Si Kα1), and from which the minimum silicon content was estimated and found to be of the order of 9 % (Table 1).

To build our arguments, let us first consider the magnetic property of pristine iron silicide. In thin films, $Fe_3Si$ is known to be ferromagnetic and to exhibit saturation magnetization as high as 140 emu/g [84]. However, it is also known that FeSi and $Fe_2Si$ have a weak ferromagnetic state and no-magnetisation, respectively [85,86]. The comparison of films and nanoparticles properties is always the most intuitive, even if it can often be misleading. The nPs are indeed very sensitive to the preparation method, size and coating. To focus on pristine silicide, we note that for instance as-prepared $Fe_3Si$ nPs with a size of about 6–9 nm obtained by pyrolysis were shown to be superparamagnetic room temperature with an $M_s$ of ~ 10 emu/g [87], and 10 nm large $Fe_3Si$ nPs obtained by colloidal pathway were also superparamagnetic and displayed an $M_s$ of ~ 60 emu/g [88]. In contrast, 150 nm large $Fe_2Si$ nanocubes presented photo-activated weaker $M_s$ of ~ 15 emu/g but were nonetheless associated with ferromagnetism at room temperature [89]. Overall the low $M_s$ value observed for the present annealed fct-FePt, when compared with the highest values reported for fct-FePt in the literature, is consistent with the formation of silicide. Evidently, the annealing is unlikely to form $Fe_xSi_y$ materials but instead to lead to $Fe_xSi_yPt$ type of alloy. As the iron, platinum and silicon atoms mix at the annealing temperature the resulting material magnetic properties should be expected to strongly deviate from fct-FePt, for which the magnetic properties originate from spin-orbit coupling and hybridisation between Fe 3d and Pt 5d states [12]. These will be disturbed by the presence of the Si atoms. Reports on $Fe_xSi_{1-x}Pt$ are scarce in the literature whether on films or on nanoparticles. The formation of silicide was demonstrated when annealing FePt films on-top of a Si substrate, where the mechanism was shown to involve the metal atoms diffusion through the native $SiO_2$ oxide layer and was concomitant with a sharp coercivity decrease [53]. Similarly, earlier works investigating the effect of annealing FePt nPs on a silicon wafer showed that a short FePt annealing resulted in a larger coercivity below the blocking temperature and an increased fct ordered fraction; whereas further annealing led to a drop of both coercivity and magnetisation [31]. This was rationalised in terms of silicon atoms diffusing from the contact point between the nPs and the substrate as illustrated in Figure 6a. The reduced magnetisation observed in these last two works is consistent with our observations. In the present study, the FePt nPs were coated with a silica shell through a process relying on a strong base, then annealed at high temperature. Under the present conditions, such a preparation pathway leads all around the nPs to silicides, which remains even after chemical etching as shown in

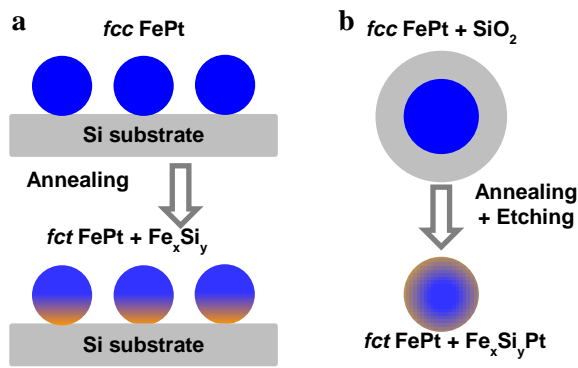

**Figure 6.** Schematic drawing of silicides formation on FePt nPs on a Si wafer (**a**) and encapsulated in a silica shell (**b**).

Figure 6b. It results in softer magnetic properties than pristine *fct*-FePt, with a strong deviation of the blocking temperature, magnetisation at saturation and magnetocrystalline anisotropy when compared to pristine *fct*-FePt. The silicide formation also explains that the *fcc*-FePt-silica nPs present the highest magnetocrystalline anisotropy of the set. In addition, it could be surprising at first sight that *fcc*-FePt-silica nPs display the highest $K_u$ and $M_s$ values, but the lowest blocking temperature. However, it could be explained by the slight increase of the volume of the particles after annealing even though we note that quantifying this by calculating the products $M_s K_u V$ is prevented by the uncertainty regarding, which size (TEM, XRD, composition) to use.

Building on the literature, the nPs interface, including the silicide layer, can be rationalised with a magnetic dead-layer which thickness can be assessed with the following equation [2,57,58,90]:

$$M_s = M_{sb} \cdot \left(1 - \frac{6 t_{dl}}{D}\right) \quad \text{eq. (5)}$$

with $M_s$ and $M_{sb}$ the magnetization at saturation of the nPs and of the bulk (~ 1140 emu/g)[58,91], respectively, $t_{dl}$ the thickness of the dead-layer and $D$ the nP size. To present a parameter independent of whether the nP size was determined by TEM or XRD, the percentage of the size corresponding to the magnetic dead-layer thickness is presented in Table 2. As expected from the magnetization at saturation measurements, the percentage of the nP size associated with the dead layer is shown to increase as the surface of the nPs was processed, reaching about 30 % with the silica layer and up to ~33 % for the *fct*-FePt nPs containing the silicide layer.

In views of the weak magnetisation properties (magnetisation at saturation, blocking temperature) there is no doubt that the side-effect of using a silica shell can in some instances prevent harvesting of the full benefit of the magnetic properties of *fct*-FePt nPs. However having identified the origin of the issue also points towards alternative new strategies to overcome it. Promising ways to address both sintering and alloy formation could rely on the design of buffer layers between the FePt nPs and the silica shell. These could include magnesium oxide, carbon or even FePtN buffer shells or matrix [24,25,92]. Alternatively, low temperature and high pressure approaches might be appropriate to convert superparamagnetic nPs self-organised on a substrate into a ferromagnetic nP thin film [93].

### IV. Conclusion

Whilst it was confirmed that silica shells prevent efficiently sintering during the nP annealing, and that *fcc*-to-*fct* conversion occurred, the interfacial silicide formation at the surface of the magnetic nPs prevented the full realisation of the stronger magnetic properties of the obtained tetragonal crystalline phase. The larger coercivity observed after thermal annealing contrasted with both the lower than expected magnetocrystalline anisotropy and magnetisation at saturation, along with the only partial recovery of the blocking temperature. Nonetheless, precisely because the main thrust of this work is to indicate the effects of silicide formation, it is worth recording its impacts and keep in mind that the observation of a tetragonal FePt phase does not alone determine the magnetic properties of nPs. As a consequence, the fact that the silicide alters qualitatively both $M_s$ and $K_u$ is an illustration of the consistency of the data and interpretation we present. Having identified the origin of the limitations associated with silica coating and dissolving, this report shed lights on poorly investigated and understood interfacial phenomena occurring during thermal annealing of nanoparticles particles. It also strengthens the interest of developing new strategies to prevent atomic diffusion and in the present case to avoid silicide formation. Promising ways to overcome this difficulty could rely on the design of buffer layers between the FePt nPs and the silica shell.

### V. Acknowledgments

The authors would like to thank the *James and Enid Nicol Trust* for funding SC's studentship, the Canon Foundation in Europe for supporting PA's visits at the RIKEN and his Fellowship, and the Ministry of Science, ICT & Future Planning, Korea (201000453, 2015001948, 2014M3A6B3063706) for hosting PA's visits during the final write-up and submission stages of the manuscript.